\documentclass[12pt]{article}

\usepackage{color}
\usepackage{amsmath,amssymb}
\usepackage{graphicx}
\usepackage{epstopdf}
\usepackage{dcolumn}
\usepackage{bm}
\usepackage[numbers,super,comma,sort&compress]{natbib}

\newcommand{\Fig}[1]{Fig.\,\ref{#1}}
\newcommand{\Eq}[1]{Eq.\thinspace (\ref{#1})}

\newcommand{\MatrK}{\textbf {K}}

\def\bR{\mbox{\boldmath $R$}}
\def\bRz{\mbox{{\boldmath $R$}$_0$}}
\def\bRg{\mbox{\boldmath $R_g$}}

\def\bfe{\mbox{\boldmath $e$}}
\def\bs{\mbox{\boldmath $s$}}
\def\bQ{\mbox{\boldmath $Q$}}
\def\bQs{\mbox{\fontsize{9}{11}\selectfont\boldmath $Q$}}
\def\bqs{\mbox{\fontsize{9}{11}\selectfont\boldmath $q$}}

\def\bq{\mbox{\boldmath $q$}}
\def\bP{\mbox{\boldmath $P$}}

\def\bff{\mbox{\boldmath $f$}}

\def\nl{\nonumber \\}
\def\be{\begin{equation}}
\def\ee{\end{equation}}
\def\bea{\begin{eqnarray}}
\def\eea{\end{eqnarray}}

\def\ie{i.e.}

\def\h3o2{H$_{3}$O$_{2}^-$}

\def\bq{\mbox{\boldmath $q$}}

\title{Path Integral Approach to the Calculation of Reaction Rates for a Reaction Coordinate Coupled to a Dual Harmonic Bath}
\author{Yonggang Yang\thanks{State Key Laboratory of Quantum Optics and Quantum Optics Devices, Shanxi University, Taiyuan 030006, P.R. China}, 
Oliver K\"{u}hn\thanks{Institut f\"ur Physik, Universit\"at Rostock, D-18051 Rostock, Germany}}

\begin{document}

\maketitle

\begin{abstract}
We present a new method for the numerical calculation of canonical reaction rate constants in complex molecular systems, which is based on a path integral formulation of the flux-flux correlation function. Central is the partitioning of the total system into a relevant part coupled to a dual bath. The latter consists of two parts: First, a set of strongly coupled harmonic modes, describing, for example, intramolecular degrees of freedom. They are treated on the basis of a reaction surface Hamiltonian approach. Second, a set of bath modes mimicking an unspecific environment modeled by means of a continuous spectral density. After deriving a set of general equations expressing the canonical rate constant in terms of appropriate influence functionals, several approximations are introduced to provide an efficient numerical implementation. Results for an initial application 
to the H-transfer in 6-Aminofulvene-1-aldimine are discussed.
\end{abstract}

  \makeatletter
  \renewcommand\@biblabel[1]{#1.}
  \makeatother

\bibliographystyle{apsrev}

\renewcommand{\baselinestretch}{1.1}
\normalsize

\clearpage

\section*{\sffamily \Large INTRODUCTION}
\label{sec:intro}
Sophisticated experimental methods nowadays provide a rather
detailed insight into molecular dynamics, unraveling the importance
of quantum effects even in rather complex systems at room
temperature \cite{may11}. This provides a challenge to theory since a fully
quantum mechanical description of condensed phase dynamics remains
out of reach and therefore approximate methods have to be developed.
Among the oldest problem is that of chemical reaction rates which in
fact gives a straightforward means for identifying quantum tunneling
in terms of the non-Arrhenius behavior.  Different methods have been
developed to account for quantum effects in rate calculations (see,
e.g., reviews in Refs. \citenum{pu06:3140,braun-sand06_1171} which set the focus on
enzyme reactions or the recent developments in Ref. \citenum{wang06:174502}).
A rigorous formulation of quantum mechanical rate
constants can be given on the basis of the path integral approach
\cite{miller83:4889,jaquet85:2139,voth02:8365}, see also the related
instanton type approaches, e.g. in Refs. \citenum{miller03_1329,yamamoto04:3086,
smedarchina07:314}.

Canonical rate constant are commonly calculated
using the  flux-flux correlation approach \cite{miller83:4889},
which requires a path integral propagation in complex time. Here, a
breakthrough in numerical efficiency has been the quasi-adiabatic
propagator (QUAPI) approach developed by Makri and coworkers
\cite{makri92:435,topaler93:285,topaler94:7500,topaler96:4430}. For
the case of a generic system-bath model the QUAPI
approach is based on a propagator splitting where the
quasi-adiabatic path along which  the bath oscillators are at their
minimum position along the reaction path serves as the reference. In
the context of rate calculations it has been applied to the
situation of a double well, bilinearly coupled to a harmonic bath
\cite{topaler93:285,topaler94:7500,forsythe99:103}, and to
electronically nonadiabatic reactions in Ref. \citenum{topaler96:4430}.
In another application Makri and Forsythe \cite{forsythe98:6819}
used the all Cartesian reaction surface Hamiltonian approach
\cite{ruf88:1523} to determine a system-bath Hamiltonian for  H
diffusion in a silicon lattice. Employing a flexible bath reference
for the Si environment led directly to the form of the Hamiltonian
used in the QUAPI method.  However, to account for the
two-dimensional motion of H an  effective one-dimensional
Hamiltonian had been used which was supplemented by an orthogonal
harmonic mode with position-dependent frequency. The Si lattice bath
modes were treated at the transition state geometry, i.e. mode-mode
coupling and coordinate-dependence of the Hessian were neglected. In
a subsequent publication, the issue of coupled bath modes has been
addressed for a generic system \cite{shao99:269}.

In the present work we consider the more general  situation, where a large
amplitude reaction coordinate belongs to some  polyatomic molecule,
which is further embedded in some environment such as a solvent or a solid state matrix. The term polyatomic
molecule is assumed to include situations with strongly coupled
solvation shells.  This setup will be termed system coupled to a
\emph{dual bath}. For the case mentioned the distinction between
intra- and intermolecular baths is motivated by the following
observation: Quite often one faces a situation where the
(intramolecular) reaction coordinate is strongly coupled to specific
intramolecular modes with an interaction potential that is not of
the standard bilinear form. This coupling can be well-described by a
reaction surface Hamiltonian, which in principle is amenable to an
ab initio treatment. For the surrounding solvent this level of
sophistication is often not necessary as the  spectral
densities associated with the coupling are broad and featureless. This suggests a treatment in terms of empirical models or  classical calculations of respective correlation functions, e.g., on the basis
of  molecular mechanics force fields \cite{may11}.

The goal of the present paper is to develop a path integral
expression for  the calculation of canonical rate constants for a
reaction coordinate coupled to a dual bath. This approach is
applied to the case of H-transfer in 6-Aminofulvene-1-aldimine
embedded in some model environment. Although our results for this
case are of preliminary character, this reaction in principle shows
some interesting effects. It was investigated in detail using NMR
spectroscopy by Limbach and coworkers \cite{lopez-del-amo08:8620}.
The temperature-dependent rate was found to be sensitive to the phase of the surrounding medium, which was either
amorphous or crystalline. In general
the reaction in the amorphous phase proceeds faster and the observed
kinetic isotope effect (KIE) becomes temperature independent
for low temperatures; at $T=$298 K it was $k^{\rm H}/k^{\rm D}=4$.
In contrast for the   crystalline environment the KIE was
temperature dependent  throughout the measured range, which did not
include tunneling regime; at $T=$298 K it was $k^{\rm H}/k^{\rm
D}=9$. The analysis of the experimental data was performed using the
Bell-Limbach model \cite{limbach04:17}. This model introduces the
reorganization energy for H-bond compression which is  necessary for
tunneling to occur from the  intrinsic barrier for the transfer in
the compressed state assuming a two step process. Further, a heavy
atom mass effect is assumed for the transferred particle. Based
on this model the effective barrier was estimated to be 3 kcal/mol
and 1.9 kcal/mol for the crystalline and amorphous phase, respectively.
In both cases the reorganization energy amounted to 0.5 kcal/mol and
the mass effect was found to be 1 a.m.u. Accounting for zero-point
energy effects yielded an effective barrier for D transfer of 1.2
kcal/mol and 0.7 kcal/mol for the crystalline and amorphous phase,
respectively. Although this model is of empirical character it shows
the importance of the specific coupling to bond-compressing
intramolecular modes as well as the influence of the environment on
the reaction rates, thus illustrating the essence of the present
dual bath approach.

In the following we will start by introducing the system-bath
Hamiltonian; a brief summary of the
derivation of the intramolecular reaction surface Hamilton is given
in the Appendix. Afterwards the path-integral expression of the canonical rate
constant will be derived and some
approximations simplifying the numerical treatment will be  introduced.
Subsequently, the application to the H/D-transfer in 6-Aminofulvene-1-aldimine
is discussed and we conclude with a summary.

\section*{\sffamily \Large THEORY} 
\subsection*{\sffamily \large System-Dual Bath Hamiltonian}
\label{sec:sb}
Large amplitude motions of certain coordinates of a polyatomic molecule
embedded in some environment will be described as a relevant
low-dimensional system coordinate, $s$, coupled to a dual
bath, where the intramolecular and environmental degrees of freedom
(DOFs) are denoted $\bQ$ and $\bq$, respectively. In the Appendix we
give a brief account on the derivation of a reaction surface
Hamiltonian for the intramolecular problem, \cite{giese06:211}
specified to the case of a linear reaction path
\cite{miller88:6298}. The resulting intramolecular Hamiltonian  can
be written as the sum of the (one-dimensional) reaction coordinate
part (we use mass-weighted coordinates and atomic units throughout)
\be\label{eq:H0}
H_0= -\frac{1}{2}\frac{\partial^2}{\partial s^2}  + V_0(s) \, ,
\ee
an intramolecular bath part%
\be
\label{eq:H1}
H_1 = \frac{1}{2}\sum_{k}\Big[ -  \frac{\partial^2}{\partial Q_{k}^2} + \omega_k^2 Q_k^2 \Big] \, ,
\ee
and a coupling part
\be
V_{1}(s,\bQ) = -\sum_k f_k\left(s\right) Q_{k}
+\frac{1}{2} \sum_{k,k'} K_{kk'}\left(s\right) Q_{k}Q_{k'} -\frac{1}{2}\sum_{k}\omega_k^2 Q_{k}^2
\ee
Here, $\bff(s)$  is the vector of forces exerted on the oscillators
(Eq. (\ref{pot2})), $\MatrK(s)$ is the reaction coordinate dependent
force constant matrix, and $\omega_k^2=K_{kk}(s_{\rm ref})$ is the
frequency of the $k$th bath mode at some reference value of the reaction coordinate.

The coupling of the reaction coordinate and the intramolecular DOFs to the harmonic bath of the environment,
\begin{equation}
H_2=\frac{1}{2}\sum_{\alpha}\Big[ -\frac{\partial^2}{\partial
q_{\alpha}^2} +\omega_{\alpha}^2 q_{\alpha}^2 \Big] \, ,
\end{equation}
will be assumed to include the lowest-order terms of a Taylor expansion with respect to $\bq$, i.e.
\begin{equation}
\label{ }
V_2(s,\bQ,\bq)= \sum_{\alpha} d_{\alpha}(s) q_{\alpha}
+\sum_{\alpha,k} C_{\alpha,k}(s) Q_k q_{\alpha} \, .
\end{equation}
Here, where $d_{\alpha}(s)$ and  $c_{\alpha,k}(s)$ are some coupling functions to be specified for the system at hand.
Thus the total Hamiltonian is given as
\be
\label{eq:htot}
H=H_0(s)+H_1(\bQ)+H_2(\bq)+V_1(s,\bQ) +V_2(s,\bQ,\bq) \, .
\ee
%
\subsection*{\sffamily \large Canonical Quantum Reaction Rate}\label{sec:rate}
%
We will use the  flux-flux correlation function expression of the
reaction rate between reactant and product, $k_{RP}$, due to Miller and coworkers \cite{miller83:4889}
\be 
\label{eq:rate}
k_{RP}=\frac{1}{Z}\int_{0}^{\infty} C_{\rm f}(t)dt  \, ,
\ee
where
\be 
C_{\rm f}(t)=\mathrm{Tr}\left\{F e^{iHt_{\rm c}^*} F
e^{-iHt_{\rm c}}\right\}
 \ee
is the autocorrelation function of the symmetrized flux operator
specified here to the case of a one-dimensional reaction coordinate
$s$ with the dividing surface at $s=0$, $F=\frac{1}{2}\left(p_s
\delta(s)+\delta(s) p_s\right)$, and  $Z=\mathrm{Tr}\left(-\beta
H_{\rm R}\right)$ is the canonical partition function of suitably defined
 reactant Hamiltonian $H_{\rm R}$.
The complex time, $t_{\rm c}=t-i \beta /2$, is due to the
combination of the time evolution operator and the Boltzmann operator and
$\beta=1/k_{\rm B} T$.

The flux autocorrelation function can be calculated by approximating
the momentum operator in the vicinity of the dividing surface by a
finite difference expression with increment $\Delta s$\cite{topaler93:285}
\be\label{deltas}
C_{\rm f}(t)=\frac{1}{2\Delta s^2}
\mathrm{Re}\left[K(\Delta s,\Delta s,0,0,t_{\rm c})-K(0,\Delta
s,0,\Delta s,t_{\rm c})\right],
\ee
where
\begin{equation}
\label{eq:prop}
K(s,s',s'',s''',t_{\rm c}) = \int_{-\infty}^{\infty}d\bQ
d\bq \langle \bQ |\langle \bq | \langle s''' | e^{iHt_{\rm c}^{*}} |
s'' \rangle \langle s' | e^{-iHt_{\rm c}} | s \rangle | \bq \rangle|
\bQ \rangle \, .
\end{equation}
The elementary propagators in this expression can be evaluated
using the path integral technique, i.e.  dividing the complex time
$t_{\rm c}$ into $N$ slices. This yields \cite{forsythe98:6819}
\bea\label{pathintgrl}
&&K(s_1,s_{N+1},s_{N+2},s_{2N+2},t_{\rm c})
\nl &=&\int_{-\infty}^{\infty} \cdots \int_{-\infty}^{\infty} d\bQ
d\bq ds_2 \cdots ds_N ds_{N+3} \cdots ds_{2N+1} \nl &&\times\langle
\bQ | \langle \bq | \prod_{n=2N+1}^{N+2} \langle s_{n+1} |
e^{-iH\delta_n} | s_{n} \rangle \prod_{n=N}^{1} \langle s_{n+1} |
e^{-iH\delta_n} | s_{n} \rangle| \bq \rangle| \bQ \rangle \nl
&=&\int_{-\infty}^{\infty} \cdots \int_{-\infty}^{\infty} ds_2
\cdots ds_N ds_{N+3} \cdots ds_{2N+1} F_{\rm
infl}(s_1,s_2,\cdots,s_{2N+2},t_c) \nl &&\times \prod_{n=2N+1}^{N+2}
\langle s_{n+1} | e^{-iH_0(s)\delta_n} | s_{n} \rangle
\prod_{n=N}^{1} \langle s_{n+1} | e^{-iH_0(s)\delta_n} | s_{n}
\rangle \, ,
\eea
where the time steps $\delta_n$ are defined as follows:
\bea
\delta_{2N+2}&=&\delta_{N+2}=\frac{-t_c^*}{2N} \nl
\delta_{n}&=&\frac{-t_c^*}{N},n=N+3,\cdots,2N+1 \nl
\delta_{N+1}&=&\delta_{1}=\frac{t_c}{2N} \nl
\delta_{n}&=&\frac{t_c}{N},n=2,\cdots,N.
\eea
In Eq. (\ref{pathintgrl}) the influence  functional is defined as
\begin{equation}
\label{ } F_{\rm infl}\left(\{s_n\}\right)=\int_{-\infty}^{\infty}
d\bq d\bQ \langle \bq | \langle\bQ | \prod_{n=2N+2}^{1} e^{-i\left[
H_1\left(\bQs\right)+V_1\left(s_n,\bQs\right)+H_2\left(\bqs
\right)+V_2\left(s_n,\bQs, \bqs \right) \right] \delta_n} |
\bQ\rangle | \bq\rangle \, ,
\end{equation}
where $\{s_n\}$ denotes a specified path realization.
With the help of the exact propagator for harmonic oscillators
\cite{feynman65} one can obtain the following result, e.g., for the
intramolecular bath part
\bea\label{propg}
\langle \bQ_{n+1} | e^{-i(H_1\left(\bQs\right)+V_1\left(s_n,\bQs\right))\delta_n} | \bQ_{n} \rangle
=\exp\left\{-i V_1(s_n,\bQ_n)\delta_n\right\}
\prod_k\sqrt{\frac{\omega_k}{2\pi i\sin(\omega_k \delta_n)}} \nl
\times \exp\left\{ \sum_k i \omega_k\left[
\cot(\omega_k \delta_n) Q_{nk}^2
-\frac{Q_{n+1,k}Q_{nk}}{\sin(\omega_k \delta_n)} \right]\right\} \, .
\eea
Note that $V_1(s_n,\bQ_n)$ not only contains the force on the
oscillator coordinates but also the mode-mode coupling and the
change of the diagonal elements of the force constant matrix with
respect to the chosen reference value of the reaction coordinate.
Actually the  choice of the latter does not play an important role,
if one  assumes that $\delta_n$ is chosen to be sufficiently small.

Using a similar expression for the environmental part of the
Hamiltonian, we arrive at the following influence functional
\be\label{newinflu}
F_{\rm infl}\left(\{s_n\}\right)=F_q F_Q
\int_{-\infty}^{\infty} d\bQ_1 \ldots d\bQ_{2N+2}d\bq_1 \ldots
d\bq_{2N+2} \exp\{g(\{s_n\},\bQ,\bq)\} \ee \bea
g(\{s_n\},\bQ,\bq)&=&\sum_{nk} i \omega_k\left[\left( \cot(\omega_k
\delta_n)+\frac{(\omega_k\delta_n)}{2} \right) Q_{nk}^2 -\frac{
Q_{n+1,k}Q_{nk}}{\sin(\omega_k\delta_n)} \right] \nl &&+ i\sum_{nk}
\delta_n f_k(s_n) Q_{nk}
-\frac{i}{2}\sum_{nkk'}Q_{nk} K_{kk'}(s_n)Q_{nk'} \nl &&+\sum_{n
\alpha} \frac{i \omega_{\alpha}}{\sin(\omega_{ \alpha}\delta_n)}
\left[ \cos(\omega_{ \alpha} \delta_n) q_{n \alpha}^2 -
q_{n+1,\alpha}q_{n \alpha} \right] \nl &&- i\sum_{n \alpha}\delta_n
d_{\alpha}(s_n) q_{n \alpha} -i\sum_{nk \alpha}\delta_n C_{\alpha
k}(s_n)Q_{nk}q_{n \alpha},
\eea
where
\begin{equation}
\label{eq:Fq}
F_q=\prod_{n \alpha}\sqrt{\frac{\omega_{\alpha}}{2\pi i\sin(\omega_{\alpha}
\delta_n)}}
\end{equation}
and
\begin{equation}
\label{eq:FQ}
F_Q=\prod_{nk}\sqrt{\frac{\omega_k}{2\pi
i\sin(\omega_k \delta_n)}}
\end{equation}
are path independent prefactors.

The partition function can be calculated following the same lines
\bea
Z&=&\int_{-\infty}^{\infty} \cdots \int_{-\infty}^{\infty} ds_1 ds_2 \cdots ds_{N_{\beta}}
F_{\beta}(s_1,s_2,\cdots,s_{N_{\beta}}) \nl
&\times& \langle s_{1} | e^{-iH_0(s)\delta_{\beta}} | s_{N_{\beta}} \rangle
\prod_{n=N_{\beta}-1}^{1} \langle s_{n+1} | e^{-iH_0(s)\delta_{\beta}} | s_{n} \rangle
\eea
with the influence functional
\begin{equation}
\label{ } F_{\beta}\left(\{s_n\}\right)=\int_{-\infty}^{\infty} d\bq
d\bQ \langle \bq | \langle\bQ | \prod_{n=N_{\beta}}^{1} e^{-i\left[
H_1\left(\bQs\right)+V_1\left(s_n,\bQs\right)+H_2\left(\bqs
\right)+V_2\left(s_n,\bQs, \bqs \right) \right] \delta_\beta} |
\bQ\rangle | \bq\rangle \, .
\end{equation}
Here $\delta_{\beta}=-{i\beta}/{N_{\beta}}$ and $N_{\beta}=2N$
is the number of time slices for the imaginary time $-i \beta$.

In a next step we need to evaluate the integrals in \Eq{newinflu} which are of the following
complex-coefficient Gaussian type
\be\label{gaussint} \int_{-\infty}^{\infty} \cdots
\int_{-\infty}^{\infty} dx_1 dx_2 \cdots dx_N \exp\{
-\sum_{mn}A_{mn}x_m x_n +i\sum_{mn}B_{mn}x_m x_n
+\sum_{n}W_{n}x_n\}, \ee where both $\mathbf{A}$ and $\mathbf{B}$
are real symmetric matrices. For any physically meaningful case
 the matrix $\mathbf{A}$ is positive-definite and the integration will converge.
 Moreover, it is possible to find one invertible real matrix, $\mathbf{U}_{\rm c}$,
 to congruently diagonalize $\mathbf{A}$ and $\mathbf{B}$
simultaneously. If $\mathbf{U_1}$ and $\mathbf{U_2}$ are orthogonal matrices such that
\bea\label{congrutrans}
\mathbf{U_1}^{\rm T} \mathbf{A} \mathbf{U_1}&=&\mathbf{a}
\equiv \mathrm{diag} \{a_1,a_2,\cdots,a_N\} \nl
\mathbf{U_2}^{\rm T} \mathbf{a}^{-\frac{1}{2}} \mathbf{U_1}^ {\rm T}
\mathbf{B}
\mathbf{U_1} \mathbf{a}^{-\frac{1}{2}} \mathbf{U_2}&=&\mathbf{b}
\equiv \mathrm{diag} \{b_1,b_2,\cdots,b_N\} \,
\eea
and the matrix $\mathbf{A}$ is positive-definite such that all
eigenvalues $\{a_n\}$ are positive, one can define the
transformation matrix 
\be \mathbf{U}_{\rm c}=\mathbf{U_1}
\mathbf{a}^{-\frac{1}{2}} \mathbf{U_2}\, 
\ee 
which transforms $\mathbf{U}_{\rm c}^{\rm T} \mathbf{A} \mathbf{U}_{\rm
c}=\mathbf{1}$ and $\mathbf{U}_{\rm c}^{\rm T} \mathbf{B}
\mathbf{U}_{\rm c}=\mathbf{b}$.  Using the new variables \{$y_n$\}
defined by $y_n=\sum_m (\mathbf{U}_{\rm c}^{-1})_{nm}x_m$ the
integration in \Eq{gaussint} can be performed analytically to give \bea
\label{eq:inte} &&\int_{-\infty}^{\infty} \cdots
\int_{-\infty}^{\infty} dx_1 dx_2 \cdots dx_N \exp\{
-\sum_{mn}A_{mn}x_m x_n +i\sum_{mn}B_{mn}x_m x_n +\sum_{n}W_{n}x_n\}
\nl &=&|\mathrm{Det}(\mathbf{U}_{\rm c}^{-1})|
\int_{-\infty}^{\infty} \cdots \int_{-\infty}^{\infty} dy_1 dy_2
\cdots dy_N \exp\{ -\sum_{n}(1-ib_n)y_n^2+\sum_{n}w_{n}y_n\} \nl
&=&\prod_{n}\left[ \sqrt{\frac{\pi}{a_n}} \sqrt{\frac{1}{1-ib_n}}
\exp\left(\frac{w_n^2}{4(1-ib_n)}\right) \right], \eea where
$w_n=\sum_m (\mathbf{U}_{\rm c})_{nm}W_m$. Here and in the following
the square root of a complex number means its principal value, i.e.,
the  non-negative real part.

Using this method it is at least in principle  possible to solve
\Eq{newinflu}. However, in practice this would imply to  numerically
diagonalize a large matrix for \emph{each} specified path. In order
to simplify matters we reconsider the  environmental bath part.
Here, the  quadratic coefficients of the bath oscillators are path
independent and assumed to be uncorrelated between each other. Based
on above mentioned procedure we can find a frequency-dependent real
invertible matrix $\textbf{U}_q(\omega)$ to congruently diagonalize
each bath mode
\be\label{diagbath}
\tilde{q}_{n}=\sum_{n'}[\textbf{U}^{-1}_q(\omega)]_{nn'}q_{n'} \ee
such that \be \sum_{n} \frac{i \omega}{\sin(\omega\delta_n)}
\left[\cos(\omega\delta_n)q_{n}^2 - q_{n+1}q_{n} \right] =-\sum_{n}
(1-i b^q_n(\omega))\tilde{q}_{n}^2 \, ,
\ee
where  the $\{b^q_n(\omega)\}$ (and $\{a^q_n(\omega)\}$ which will
appear below) are the eigenvalues from  diagonalizing the
corresponding coefficients matrix according the procedure introduced
in \Eq{congrutrans}. Using the new variables $\tilde{q}_{n
\alpha}=\sum_{n'}[\textbf{U}^{-1}_q(\omega_{\alpha})]_{nn'}q_{n'
\alpha}$ the integration over $\{\tilde{q}_{n \alpha}\}$ can be
performed analytically. The final result for influence functional is
given by
\bea\label{finalinflu}
F_{\rm infl}\left(\{s_n\}\right)&=&F_q F_Q
\tilde{F_q} \int_{-\infty}^{\infty} \cdots \int_{-\infty}^{\infty}
d\bQ_1d\bQ_2 \cdots d\bQ_N\exp\{g(\{s_n\},\bQ)\} \nl
g(\{s_n\},\bQ)&=&\sum_{nk} i \omega_k \left[\left( \cot(\omega_k
\delta_n)+\frac{(\omega_k\delta_n)}{2} \right) Q_{nk}^2 -\frac{
Q_{n+1,k}Q_{nk}}{\sin(\omega_k\delta_n)} \right] \nl &&+
i\sum_{nk}\delta_n f_k(s_n) Q_{nk}
-\frac{i}{2}\sum_{nkk'}Q_{nk}K_{kk'}(s_n)Q_{nk'}+\Delta(\{s_n\})
\nl &&+i\sum_{nk}\delta_n \Delta f_k(s_n) Q_{nk}
+\sum_{nkn'k'}g_{nk,n'k'}(s_n,s_{n'})Q_{nk}Q_{n'k'},
\eea
where
\begin{equation}
\label{eq:Fqt}
\tilde{F_q}=\prod_{n\alpha}\sqrt{\frac{\pi}{a^q_{n}(\omega_{\alpha})}}\sqrt{\frac{1}{(1-ib^q_{n}(\omega_{\alpha}))}}
\end{equation}
\bea
\Delta(\{s_n\})&=&\sum_{n\alpha}\frac{w_{n\alpha}^2}{4[1-ib^q_{n}(\omega_{\alpha})]}\nl
w_{n\alpha}&=&-i\sum_{n'}[\textbf{U}_q(\omega_{\alpha})]_{n'n}
\delta_{n'}d_{\alpha}(s_{n'})\nl \Delta
f_k(s_n)&=&\sum_{n'\alpha}\frac{w_{n'\alpha}u_{n'\alpha,nk}}{2[1-ib^q_{n'}(\omega_{\alpha})]}\nl
u_{n'\alpha,nk}(s_n)&=&-i[\textbf{U}_q(\omega_{\alpha})]_{nn'}\delta_{n}C_{\alpha
k}(s_n) \nl
g_{nk,n'k'}(s_n,s_{n'})&=&\sum_{n''\alpha}\frac{u_{n''\alpha,nk}u_{n''\alpha,n'k'}}{4[1-ib^q_{n''}(\omega_{\alpha})]}.
\eea

So far the $\bq$-integrations have been performed following the idea from \Eq{gaussint} to \Eq{eq:inte}. The quantities entering  \Eq{finalinflu} can be calculated readily before the $\bQ$-integrations. In a final step the procedure of \Eq{congrutrans} can be applied to the $\bQ$-integrations to 
numerically diagonalize the complex coefficient matrix in \Eq{finalinflu} for each path of the reaction coordinate. The final result for the influence functional of
the reaction coordinate plus dual bath system can then be formally written as:
\be\label{influresult} 
F_{\rm infl}\left(\{s_n\}\right)=F_q F_Q
\tilde{F_q} e^{\Delta(\{s_n\})} \prod_{nk}\left(
\sqrt{\frac{\pi}{a^Q_{nk}}} \sqrt{\frac{1}{1-ib^Q_{nk}}}
\exp\Big\{\frac{w_{nk}^2}{4(1-ib^Q_{nk})}\Big\} \right) \,. 
\ee
with the different functions defined in Eqs.(\ref{eq:Fq}),
(\ref{eq:FQ}) and (\ref{eq:Fqt}). The quantities $a_{nk}^Q$,
$b_{nk}^Q$, and $w_{nk}$ in above formal expression can be obtained
from the numerical diagonalization of the respective complex coefficient matrix.
%
\subsection*{\sffamily \large Approximations}
\label{sec:approx}
%
Depending on the system size obtaining the quantities in \Eq{influresult} by direct diagonalization might
become rather time consuming, due to those terms which depend on the system's coordinate and, therefore,
have to be evaluated for each specific path. Therefore, we will introduce certain approximations
to make the approach numerical efficient for such cases.

First, we will assume that the coupling strength between the $Q_k$ and $q_{\alpha}$ modes
does not strongly depend on $s$. Thus we ignore the $s$-dependence
of the coupling strength between $Q_k$ and $q_{\alpha}$, i.e,
$\{C_{\alpha k}\}$ are simply constants and hence
$\{g_{nk,n'k'}(s_n,s_{n'})=g_{nk,n'k'}\}$ are also constants.
Next, we  assume that not for all modes, $\{Q_k\}$, the mode mixing due to $\mathbf{ K}(s)$ shows a strong coordinate dependence.  In the following we use
$\{Q_k\}$ to denote those intramolecular DOFs, which are most strongly affected by the motion of the reaction coordinate $s$. The remaining intramolecular modes are comprised in $\{Q_\nu\}$. For the latter modes, the quadratic
coefficients will be replaced by their $s$-independent mean values
along the reaction path in \Eq{finalinflu}, i.e.,
\be
K_{\nu\nu'}(s_n) \rightarrow
\langle K_{\nu\nu'} \rangle  \equiv
\frac{1}{2L}\int_{-L}^{L}K_{\nu\nu'}(s)ds,
\ee
where $2L$ is the length of the reaction path. Under this approximation, we need
to diagonalize a large matrix just once, while for each specified path we only
need to diagonalize a much smaller matrix since only a few DOFs, $\{Q_k\}$,
are significantly coupled via $s$.

Following the idea of \Eq{congrutrans} we can find a real invertible
matrix $\mathbf{U^Q}$ which congruently diagonalizes the quadratic
coefficient matrix related only to $\{Q_{\nu}\}$ 
\bea 
&&\sum_{n \nu}
i \omega_{\nu}\left[\left(
\cot(\omega_{\nu}\delta_n)+\frac{(\omega_{\nu} \delta_n)}{2} \right)
Q_{n\nu}^2 -\frac{Q_{n+1,\nu}Q_{n\nu}}{\sin(\omega_{\nu} \delta_n)}
\right] \nl &&-\frac{i}{2}\sum_{n\nu\nu'}\delta_n \langle K_{\nu\nu'} \rangle
Q_{n\nu}Q_{n\nu'} +\sum_{n\nu,n'\nu'} g_{n\nu,n'\nu'}
Q_{n\nu}Q_{n'\nu'} \nl &=&-\sum_{n\nu} (1-ib_{n\nu})
\tilde{Q}_{n\nu}^2, 
\eea 
where
$\tilde{Q}_{n\nu}=\sum_{n'\nu'}\left(\mathbf{U^Q}\right)^{-1}_{n\nu,n'\nu'}Q_{n\nu'}$.
With the help of this transformation we can analytically integrate
over the $\{\tilde{Q}_{n\nu}\}$ part. This will further contribute a
pre-factor $\tilde{F_Q}$ and some modifications to the exponential
factor compared with \Eq{finalinflu} 
\bea\label{numinflu}
F_{\rm infl}\left(\{s_n\}\right)&=&F_q F_Q \tilde{F_q} \tilde{F_Q}
\int_{-\infty}^{\infty} \cdots \int_{-\infty}^{\infty} d\bQ_1 d\bQ_2
\cdots d\bQ_N\exp\{g(\{s_n,Q_{nk}\})\} \nl
g(\{s_n,Q_{nk}\})&=&\sum_{nk} i \omega_k\left[\left(\cot(\omega_k
\delta_n)+\frac{(\omega_k \delta_n)}{2}\right)Q_{nk}^2-\frac{
Q_{n+1,k}Q_{nk}}{\sin(\omega_k \delta_n)} \right] \nl &&+
i\sum_{nk}\delta_n f_k(s_n)
Q_{nk}-\frac{i}{2}\sum_{nkk'}Q_{nk}K_{kk'}(s_n)Q_{nk'}+\Delta(\{s_n\})
\nl &&+i\sum_{nk}\delta_n \Delta f_k(s_n) Q_{nk}
+\sum_{nkn'k'}g_{nk,n'k'}Q_{nk}Q_{n'k'}+\tilde{\Delta}(\{s_n\}) \nl
&&+i\sum_{nk}\delta_n \Delta \tilde{ f}_k(s_n) Q_{nk}
+\sum_{nkn'k'}\tilde{g}_{nk,n'k'}Q_{nk}Q_{n'k'}, 
\eea 
where
\begin{equation}
\label{eq:FQt}
\tilde{F}_Q=\prod_{n\nu}\sqrt{\frac{\pi}{a_{n\nu}}}\sqrt{\frac{1}{1-ib_{n\nu}}}
\end{equation}
%
and the additional terms caused by the reduction of DOFs are defined
as follows: \bea
\tilde{\Delta}(\{s_n\})&=&
\sum_{n\nu}\frac{w_{n\nu}^2}{4(1-ib_{n\nu})}
\nl w_{n\nu}&=&\sum_{n'\nu'}(\mathbf{U^Q})_{n'\nu',n\nu}\left[i
\delta_{n'}f_{\nu'}(s_{n'})+i \delta_{n'}\Delta
f_{\nu'}(s_{n'})\right] \nl 
\Delta\tilde{ f}_k(s_n)&=&
\sum_{n'\nu'}\frac{w_{n'\nu'}u_{nk,n'\nu'}}{2(1-ib_{n'\nu'})} \nl
u_{nk,n'\nu'}&=&
2\sum_{n''\nu}(\mathbf{U^Q})_{n''\nu,n'\nu'}g_{nk,n''\nu}-i\sum_{\nu}
(\mathbf{U^Q})_{n\nu,n'\nu'}\delta_{n}K_{k\nu}(s_n) \nl
\tilde{g}_{nk,n'k'}(s_n,s_{n'})&=&
\sum_{n''\nu}\frac{u_{nk,n''\nu}u_{n'k',n''\nu}}{4(1-ib_{n''\nu})} \, .
\eea
Similar to \Eq{influresult} the final result can be written
formally as
\be\label{influresultapp} F_{\rm infl}\left(\{s_n\}\right)=F_q F_Q
\tilde{F}_q \tilde{F}_Q e^{\Delta(\{s_n\})+\tilde{\Delta}(\{s_n\})}
\prod_{nk}\left( \sqrt{\frac{\pi}{a^Q_{nk}}}
\sqrt{\frac{1}{1-ib^Q_{nk}}}
\exp\Big\{\frac{w_{nk}^2}{4(1-ib^Q_{nk})}\Big\} \right) \nonumber.
\ee
with the different functions defined in Eqs.(\ref{eq:Fq}),
(\ref{eq:FQ}), (\ref{eq:Fqt}), and (\ref{eq:FQt}). The quantities $a_{nk}^Q$,
$b_{nk}^Q$, and $w_{nk}$ in above formal expression can be obtained
from the numerical diagonalization of the complex coefficient matrix
and the final sum is only for modes which strongly couple to $s$.
 The final numerical calculations may start from
\Eq{numinflu} which is feasible since only a very low-dimensional
matrix (according to the coordinates $\{Q_k\}$) needs to be diagonalized for each specified path.
%
\section*{\sffamily \Large APPLICATION TO THE H/D-TRANSFER IN 6-AMI\-NO\-FUL\-VENE-1-ALDIMINE}
\label{sec:appl}
\begin{figure}[t]
\centering
\includegraphics [scale=0.29] {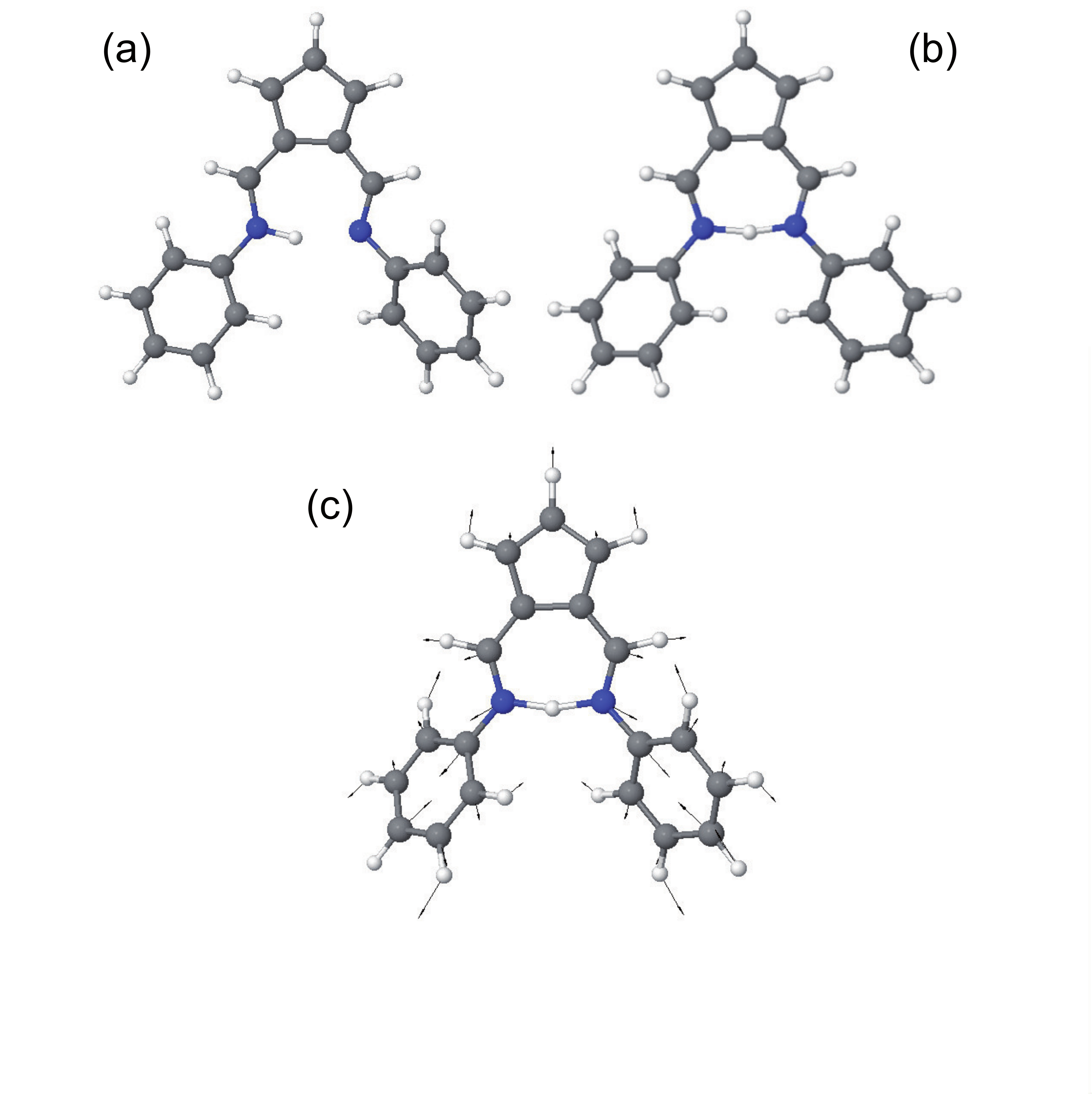}
\caption{6-Aminofulvene-1-aldimine: (a) minimum configuration, (b) transition state for the
hydrogen atom transfer, (c) Strongly coupled normal mode at the transition state having a frequency of  641 cm$^{-1}$. All results have been obtained at the B3LYP/6-31+G(d,p) level of theory.
\label{fig1}}
\end{figure}

%
In this section we present results of  a preliminary simulation based on a reaction surface model Hamiltonian describing the intramolecular H atom transfer in 6-Aminofulvene-1-aldimine. Here, our aim is not to provide a quantitative assessment of this reaction, but to illustrate the theoretical formalism presented in the previous section. The configuration of two stationary points, which have been obtained at the B3LYP/6-31+G(d,p) level of theory\cite{gaussian03} are shown in \Fig{fig1} . The minimum configuration in panel (a) corresponds to the reactant or equivalent product. The H
atom transfer process can take place from the reactant via the
transition state (panel (b)) to the product or inversely. The reaction barrier
height, as calculated by the energy difference of the minimum and the
transition state, is 3.8 kcal/mol (fully relaxed gas phase barrier). 

\begin{figure}[t]
\centering
\includegraphics [scale=0.7] {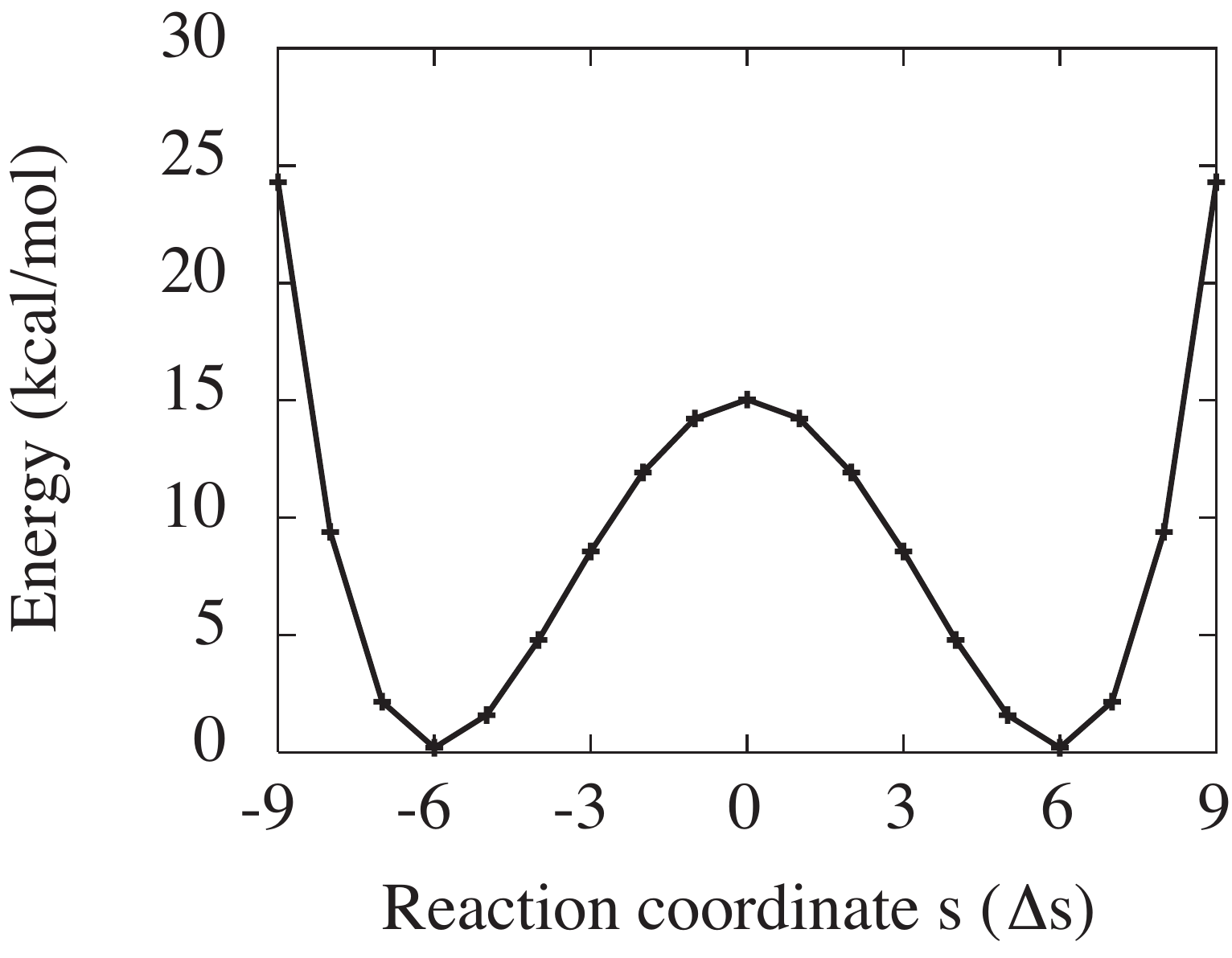}
\caption{The zeroth-order potential energy curve, $V(\bR_0)$, obtained at the B3LYP/6-31+G(d,p) level of theory for the hydrogen/deuterium atom transfer reaction
in 6-Aminofulvene-1-aldimine as shown in \Fig{fig1}. 
The curve has been generated from 19 points, symmetrically distributed with respect to the reference geometry $s_{\rm ref}=0$, which corresponds to $\bR_{\rm ref}=(\bR_{\rm prod}-\bR_{\rm reac})/2$. The reactant/product configuration is at $s=\pm 6 \Delta s$, i.e. $\Delta s= |\bR_{\rm prod}-\bR_{\rm reac}|/12$.
\label{fig2}}
\end{figure}

The unit vector which defines the linear reaction path is given by the direction pointing from the equivalent reactant to the product, \ie,
$\bfe_s=(\bR_{\rm prod}-\bR_{\rm reac})/(|\bR_{\rm prod}-\bR_{\rm reac}|)$ (cf.  \Fig{fig1}). The potential along the one-dimensional linear reaction path coordinate,
$V(\bR_0)$, is shown in \Fig{fig2}. According to the present linear reaction path, the barrier is as high as 14.85 kcal/mol. 

Within	 the harmonic approximation for the intramolecular bath modes, the energy difference with respect to the fully relaxed reaction path will be recovered by the so-called
bath reorganization energy \cite{giese06:211} (see also discussion in Refs. \citenum{giese05_054315} and \citenum{matanovic07_014309} where two and three reaction coordinates, respectively, have been used to obtain a better agreement with the fully relaxed barrier even without taking into account coupled harmonic vibrations). Furthermore, in the real system, there will be a contribution to the reorganization energy due to the interaction with the environment. In the following we do not attempt to fit the environmental contribution such as to obtain agreement with the experimental estimate by Limbach and coworkers. \cite{lopez-del-amo08:8620}

The considered molecule has 105 intramolecular vibrational degrees of freedom, whose couplings to the reaction coordinate, i.e.  $f_k(s)$, and $K_{k,k'}(s)$, can be obtained as described in the Appendix, \Eq{ke3}. For the reference $\bR_{\rm ref}$ we have chosen the point midway between reactant and product along the one-dimensional reaction path.
For the present illustration we have selected only one strongly coupled mode $Q_k$ for explicit consideration. The displacement vectors are shown in Fig. \ref{fig2}c. Apparently this mode symmetrically modifies the H-bond length and therefore modulates the reaction barrier. The remaining intramolecular modes as well as possible environmental modes are comprised into the bath $\bq$. In other words, we have simplified matters and started directly from 
\Eq{finalinflu}. 

The coupling between the environment $\bq$ and the intra-molecular DOFs, $s$ and $Q_k$, are defined as 
\bea
d_{\alpha}(s)&=&d_1 e^{-\omega_{\alpha}^{2}/d_2^2}(s+\eta s^2) \nl
C_{\alpha,k}(s)&=&C_{\alpha,k}=c_1 e^{-(\omega_k-\omega_{\alpha})^{2}/c_2^2}, 
\eea 
where $d_1$, $d_2$, $\omega_0$, $\eta$, $c_1$, and $c_2$ are parameters. The $s$-dependence of $d_{\alpha}(s)$ has been expanded to second-order and the bath
frequency dependence has been simply chosen to be of Gaussian form. The environmental modes are assumed to have uniform density of states in the region, where we take into account the coupling with the molecular DOFs.

\begin{figure}[t]
\centering
\includegraphics [scale=0.7] {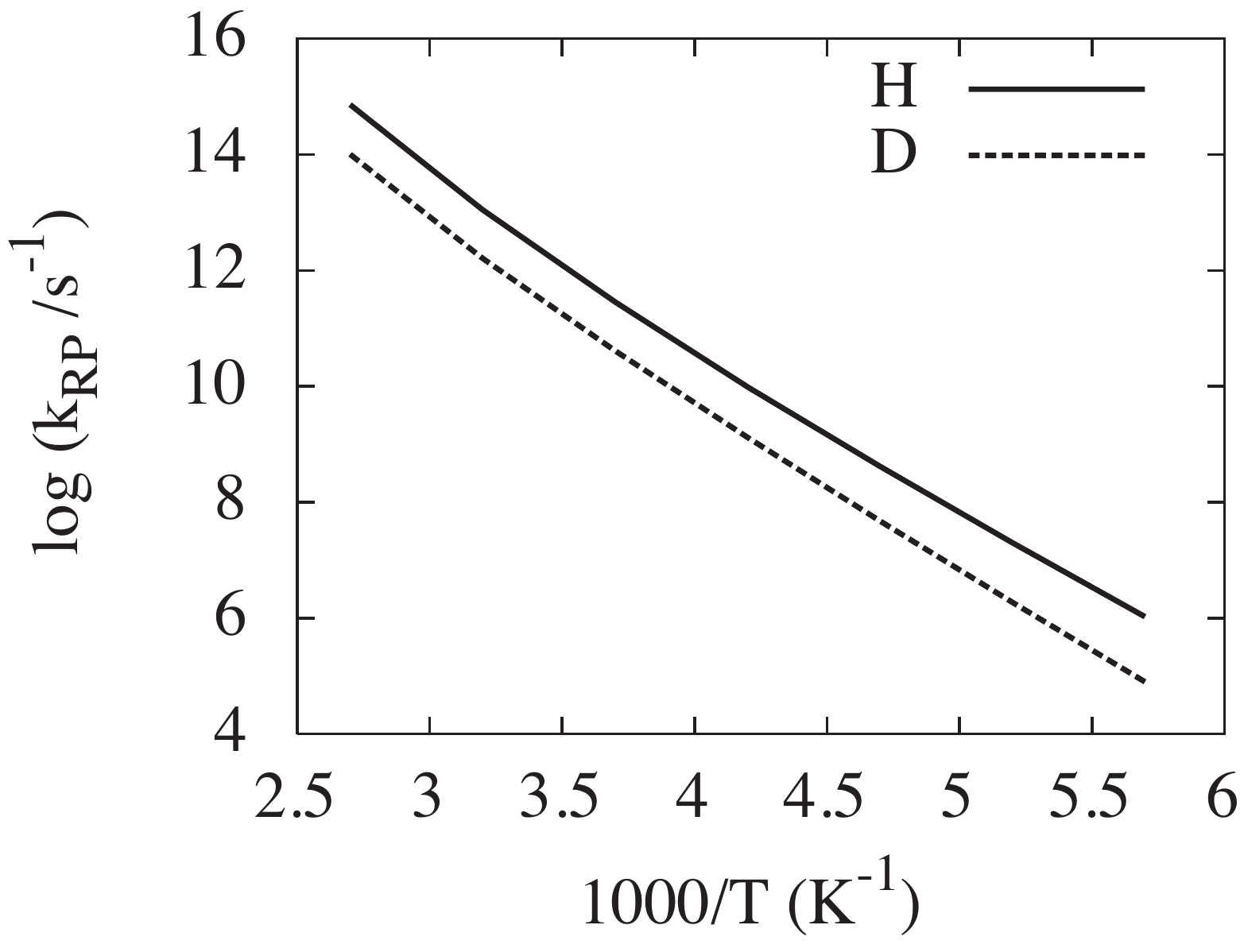}
\caption{The calculated temperature dependence of H/D transfer rate constants, \Eq{eq:rate}, in the thermal activation region based on a one-dimensional linear reaction path, $s$, coupled to one intra-molecular mode, $Q_k$, and 50 bath modes, $\bq$. In the simulation the reaction coordinate integration has been replaced by a sum over the three points, $s=-1,0,1$. For the partition function of the reactant these points were chosen as $s=-7,-6,-5$. The number of time slices has been $N=4$.
\label{fig3}}
\end{figure}
The calculated canonical rates, \Eq{eq:rate}, obtained from this  preliminary model Hamiltonian are shown in \Fig{fig3}. 
At 298 K the  KIE is $k_{RP}^H/k_{RP}^D=10$, when the following coupling parameters are used:
$c_1=d_1=$(0.628 kcal/mol)$^2$, $c_2=d_2=6.28$ kcal/mol, and $\eta=0.2 \Delta s^{-1}$.
The involved bath frequency region covers the range from 3 to 30 kcal/mol with 50 harmonic oscillators equally distributed.
Given the fact that the experimental KIE ranges between 4 and 9 and strongly depends on the phase of the environment, the present order-of-magnitude agreement is rather reasonable given the simple model for the system-bath coupling. Further, we note that the same holds true for the absolute values of the rates. However, the obtained values for the activation energies (slopes of curves in Fig. \ref{fig3}) deviate from the experimental ones. In Ref. \citenum{lopez-del-amo08:8620} it was found that the rate between the thermal activation energies between the H and the D case is about 2/3. In the present simulation the activation energies are about 3-4 times the experimental ones and the 
difference between the activation energies of different isotopomers are too small.
The latter fact is not surprising since the shapes of potential curves for hydrogen and deuterium transfers are the same and the only difference lies in the length of the step $\Delta s$ which appears in \Eq{deltas}.
The ratio for the steps is only slightly different from one, $\Delta s(H)/\Delta s(D)=0.9978$. In order to improve the description at this point, more intramolecular vibrational modes need to be taken into account. This would  lower the effective reaction barrier due to reorganization energy contributions and therefore the activation energy. On the other hand, due to the isotope dependent effective coupling, the difference between H and D activation energies would become more pronounced. 

\begin{figure}[h]
\centering
\includegraphics [scale=0.67]{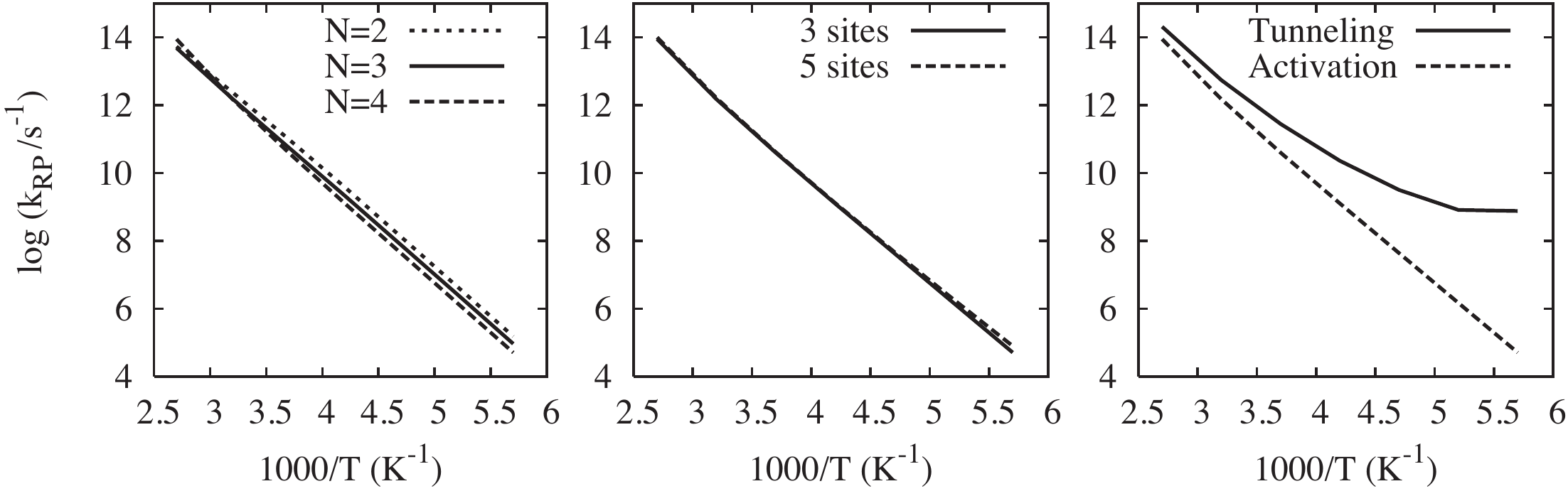}
\caption{The left and middle panels show the convergence of the
thermal activation energy (the slope) for H-transfer in the high temperature region
by only considering configurations which are important for thermal
activation, \ie, around $s=0$. In the left panel the dependence on the number of time slices $N$ is shown for $s=-1,0,1$ and in the middle panel the case  $s=-2,-1,0,1,2$ is given for $N=4$.The right panel shows
the quantum tunneling effects in the low-temperature region by
covering some configurations which are important for tunneling ($s=-2,-1,0,1,2$ (dashed curve) and $s=-5,-1,0,1,5$ (solid curve)).\label{fig4}}
\end{figure}
The numerical effort in the calculation of the propagator in \Eq{pathintgrl} by path integration depends on the number of time slices, $N$, as well as on the method for evaluating the multi-dimensional integrals of the reaction path coordinates, $s_n$. For the present application in 
\Fig{fig3} the focus has been on the thermal activation range, i.e. the high-temperature regime. This allowed us to 
simplify the rate calculation by performing the integration as a sum over three points in the vicinity of the reaction barrier, $s=-1,0,1$. In Fig. \ref{fig4} (middle panel) we show the dependence on the number of discretization points for $N=4$ time slices. Specifically, we have chosen  $s=-2,-1,0,1,2$. The ignorable
difference shows the applicability of the simplification technique, which we have adopted for the high temperature calculations. The dependence on the number of time-slices is shown in the left panel of Fig. \ref{fig4}. Clearly, the variation for the covered range, $N=2,3,4$, is rather small, justifying our choice of $N=4$ in Fig. \ref{fig3}.
Finally, we address the issue of tunneling in the right panel of Fig. \ref{fig4}. In principle, accounting for quantum tunneling at low temperatures requires to include configurations, which are located near the turning point corresponding to the  energy of the tunneling particle. In order to illustrate this point, we present results for five discretization points, i.e. $s=-2,-1,0,1,2$ (dashed curve) and $s=-5,-1,0,1,5$ (solid curve). The change of the mechanism from thermal activation to tunneling is apparent from the quite different temperature dependence of the rates. In passing we note that a systematic study of different discretizations might give an indication for those configurations that contribute to the tunneling process.
\section*{\sffamily \Large SUMMARY} 
\label{sec:concl} 
We have developed a path integral method for the determination of canonical reaction rates for the case of a reaction coordinate coupled to a dual bath. The latter is comprised of an intramolecular part, which is modeled using the reaction surface Hamiltonian approach, and an intermolecular (solvent) part. Such a partition of the interactions into different structurally motivated levels appears to be most suitable for the description of intramolecular proton or H-atom transfer reactions. The formulation benefits from the harmonic oscillator nature of the intra- and intermolecular baths in two respects: First, it enables us to perform the integration over the bath variables and second the reaction surface Hamiltonian method provides a means to determine intramolecular Hamiltonian parameters from first principles. This involves couplings between normal modes along the reaction path due to the non-diagonal Hessian matrix, which require a diagonalization for each specified path and thus substantial numerical effort. We have suggested an approximation which amounts to the replacement of the reaction coordinate dependent Hessian by its averaged value for less strongly coupled intramolecular modes. 

The initial application has been to the H/D transfer in 6-Aminofulvene-1-aldimine. Despite the various additional approximations in this application it could be shown, that our approach can give reasonable (i.e. order-of-magnitude) estimates for the reaction rates. Further work on this particular reaction shall be directed to obtain an improved description of the reaction surface as well as a realistic model for the interaction with the environment.
\subsection*{\sffamily \large ACKNOWLEDGMENTS}

We gratefully acknowledge financial support by the Deutsche Forschungsgemeinschaft through the GK 788 at the Freie Universit\"at Berlin.
Y. Yang also acknowledges financial support from National Natural Science Foundation of China under Grant No. 11004125.
%
%
\renewcommand \thesection{A}
\renewcommand{\theequation}{\Alph{section}.\arabic{equation}}
\setcounter{equation}{0}  
\numberwithin{equation}{section}
\section*{\sffamily \large APPENDIX} 
\subsection*{\sffamily \large Reaction Surface Hamiltonian}
The  reaction surface Hamiltonian combines the description of several large amplitude
coordinates \{$s_\alpha$\} coupled to many small amplitude displacements \{$Q_k$\}.
\cite{miller80:99,tew01_1958,giese05_054315} To generate
this Hamiltonian from the exact Cartesian coordinate Hamiltonian we
can directly exploit our recently developed kinetic energy
quantization method. \cite{yang08:2445}

Suppose that we have the Cartesian Hamiltonian 
\bea\label{H-tot}
H\left(\bR\right)&=&T\left(\bR\right)+V\left(\bR\right) \nl
T\left(\bR\right)&=&\frac{1}{2}\bP^2 =-\frac{1}{2}\frac{\partial
^2}{\partial \bR ^2} \, , 
\eea 
where $\bR$ is the $3N$-dimensional vector of mass-weighted Cartesian coordinates for system with $N$
atoms and $\bP=-i\partial/\partial \bR$ is the corresponding linear momentum operator. Suppose that there is a reaction surface defined by a function along the reaction coordinates $\bs$, i.e.,
 \be
\bR=\bRz\left(\bs\right) \, . 
\ee 
The potential energy function, $V\left(\bR\right)$, is expanded around the reaction surface as follows
\bea\label{pot1}
V\left(\bR\right) \approx V\left(\bRz\left(\bs\right)\right)
+\Delta\bR(\bs)^{\rm T} \frac{\partial V}{\partial
\bR}\Big|_{_{\bRz\left(\bs\right)}} +\frac{1}{2} \Delta\bR(\bs)^{\rm T}
\frac{\partial^2 V}{\partial \bR ^2}\Big|_{_{\bRz\left(\bs\right)}}
\Delta\bR(\bs) \, , 
\eea 
where $\Delta \bR\left(\bs\right)=\bR-\bRz\left(\bs\right)$. The reaction surface is defined in such a way
that the potential energy $V\left(\bR\right)$ can be approximated
by low-order  orthogonal displacements, \ie, \Eq{pot1} can be
truncated in the given form.

To obtain the reaction surface Hamiltonian we first need to define
the new coordinates, \ie, the reaction coordinates \{$s_\alpha$\}
and the orthogonal displacements \{$Q_k$\}. The former are already
defined by the reaction surface as well as the unit vectors
\{$\bfe_\alpha\left(\bs\right)$\} according to which we have the
reaction coordinate vector 
\be 
\bs=\sum_{\alpha=1}^D s_\alpha
\bfe_\alpha\left(\bs\right). 
\ee 
To get the latter we need a
projection operator to project out the reaction coordinate $\bs$ 
\be
\mathcal{P\left(\bs\right)}=1-\sum_\alpha \bfe_\alpha \bfe_\alpha^{\rm T}.
\ee 
Then we can diagonalize the projected Hessian matrix
$\mathbf{K}\left(\bs\right)$ for each point of the reaction surface
by an orthogonal transformation $\mathbf{U}_{\rm RS}\left(\bs\right)$
\be\label{trans} 
\mathbf{U}_{RS}\left(\bs\right)^\dag
\mathbf{K}\left(\bs\right) \mathbf{U}_{\rm RS}\left(\bs\right) ={\rm
diag}\{\cdots \omega_\alpha^2\left(\bs\right) \cdots
\omega_g^2\left(\bs\right) \cdots \omega_k^2\left(\bs\right)
\cdots\}, 
\ee 
where
$\mathbf{K}\left(\bs\right)=\mathcal{P\left(\bs\right)} \partial^2
V/\partial \bR ^2|_{\bRz}\mathcal{P\left(\bs\right)}$ is a real
symmetric matrix.

In total there are $D+6$ zero eigenvalues $\{\omega_\alpha^2\}$ and $\{\omega_g^2\}$ corresponding to the reaction
coordinates and six-dimensional global translation and rotation, respectively.
The orthogonal transformation matrix contains the corresponding
eigenvectors of $\mathbf{K}\left(\bs\right)$
\be
\mathbf{U}_{\rm RS}\left(\bs\right)=\left(\cdots \bfe_\alpha
\left(\bs\right) \cdots \bfe_g \left(\bs\right) \cdots \bfe_k
\left(\bs\right) \cdots\right).
\ee
The six-dimensional global translation and rotation as well as the $3N-6-D$
displacements orthogonal to the reaction surface are defined by
\bea
R_g&=&\bfe_g^{\rm T} \Delta \bR \nl
Q_k&=&\bfe_k^{\rm T} \Delta \bR.
\eea
The original $3N$-dimensional vector is now expressed with the new unit vectors
\be
\bR=\bR_{\rm ref}+\sum_{\alpha}s_{\alpha}\bfe_{\alpha}
+\sum_{g}R_g\bfe_{g}+\sum_{k}Q_k\bfe_{k},
\ee
where the reference geometry $\bR_{\rm ref}=\bRz\left(\bs=0\right)$ is
the origin of the new coordinates system.

Based on the knowledge of the new coordinates it is not difficult to find the potential energy
\be\label{pot2}
V\left(\bs,\bQ\right)=V(\bRz\left(\bs\right))
-\sum_k f_k\left(\bs\right) Q_k
+\frac{1}{2}\sum_k \omega_k\left(\bs\right)^2 Q_k^2,
\ee
where $f_k\left(\bs\right)=-\bfe_k^{\rm T}\partial V/\partial \bR|_{\bRz}$.
It is obvious that the potential energy does
not depend on $\{\bRg\}$, however, the kinetic energy operator (KEO)
does depend on $\{\bRg\}$ and normally it is not possible to 
separate them exactly. According to Ref. \citenum{yang08:2445} the following formal KEO can be obtained 
%
\be\label{ke1}
T=\frac{1}{2}\tilde{\bP}^{\dag} \frac{\partial \tilde{\bR}}{\partial \bR}
\left(\frac{\partial \tilde{\bR}}{\partial \bR}\right)^{\rm T} \tilde{\bP},
\ee
where
$\tilde{\bR}^{\rm T}=
\left(%
\begin{array}{ccc}
\bs^{\rm T} & \bRg^{\rm T} & \bQ^{\rm T} \\
\end{array}%
\right)$ 
is the full set of the new coordinates and
$\tilde{\bP}=-i\partial/\partial \tilde{\bR}$. According to
Ref. \citenum{yang08:2445} all  components of $\tilde{\bP}$ are Hermitian due to the orthogonality of
transformation except $\bP_{\bs}$. \Eq{ke1} has a fully coupled form
in case of a general reaction surface. The  factor which is responsible for complexity when it comes to a numerical implementation is that all  unit vectors depend on $\bs$, i.e., the orthogonal transformation matrix
$\mathbf{U}_{\rm RS}\left(\bs\right)$ depends on $\bs$ thus we have to
calculate the derivatives with respect to $\bs$.
\subsection*{\sffamily \large Linear Reaction Surface Hamiltonian} 
In the following we will simplify the KEO, Eq. (\ref{ke1}), by choosing a different representation in terms of
constant unit vectors that describe the reaction coordinates $\bs$ for the special case of a  
linear reaction surface. \cite{miller88:6298} With the help of certain predefined constant unit vectors $\{\bfe_{\alpha}\}$ we
can obtain the following equation for the linear reaction surface
\be
\label{eq:R0}
\bRz\left(\bs\right)=\bR_{\rm ref}+\sum_{\alpha}s_{\alpha}\bfe_{\alpha}.
\ee
The coordinate transformations are modified as follows:
\bea\label{trans2}
\bR&=&\bRz\left(\bs\right)+\sum_{k}Q_k\bfe_k
=\bR_{\rm ref}+\sum_{\alpha}s_{\alpha}\bfe_{\alpha}+\sum_{k}Q_k\bfe_k \nl
s_{\alpha}&=&\bfe_{\alpha}^{\rm T}\Delta\bR, \hspace{10 mm} Q_k=\bfe_k^{\rm T}\Delta\bR,
\eea
where $\Delta\bR=\bR-\bR_{\rm ref}$ is different from $\Delta\bR\left(\bs\right)$ in \Eq{pot1} while $\{Q_k\}$ and
$\{\bfe_k\left(\bs\right)\}$ have the same definition as in the previous section.
Note that we have combined the $\{R_g\}$ and
$\{Q_k\}$ into the same set of indexes $\{Q_k\}$ to simplify the
notation.
With the help of \Eq{ke1} and \Eq{trans2} we can
derive a simplified KEO for a linear reaction surface.
First, we calculate the elements of the Jacobi matrices
starting from \Eq{trans2}. Using the
chain rule to calculate the derivatives from \Eq{trans2} leads to the following results 
\bea 
\frac{\partial
s_{\alpha}}{\partial \bR}&=&\bfe_{\alpha}^{\rm T} \nl \frac{\partial
Q_{k}}{\partial \bR}&=&\bfe_{k}^{\rm T}
+\sum_{\alpha}\bfe_{\alpha}^{\rm T}\left(\Delta\bR^{\rm T}\frac{\partial
\bfe_{k}}{\partial s_{\alpha}}\right) \, . 
\eea 
Thus the elements for the matrix products in \Eq{ke1} can be obtained as follows
\bea 
\left(\frac{\partial
\bs}{\partial \bR} \left(\frac{\partial \bs}{\partial
\bR}\right)^{\rm T}\right)_{\alpha\beta}
&=&\bfe_{\alpha}^{\rm T}\bfe_{\beta}=\delta_{\alpha\beta} \nl
\left(\frac{\partial \bs}{\partial \bR} \left(\frac{\partial
\bQ}{\partial \bR}\right)^{\rm T}\right)_{\alpha k}
&=&\bfe_{\alpha}^{\rm T}\left(\bfe_{k} +\sum_{\beta}\bfe_{\beta}
\left(\Delta\bR^{\rm T}\frac{\partial \bfe_{k}}{\partial
s_{\beta}}\right)\right) =\Delta\bR^{\rm T}\frac{\partial
\bfe_{k}}{\partial s_{\alpha}} \nl \left(\frac{\partial
\bQ}{\partial \bR} \left(\frac{\partial \bQ}{\partial
\bR}\right)^{\rm T}\right)_{kk'} &=&\delta_{kk'}+
\sum_{\alpha}\left(\Delta\bR^{\rm T}\frac{\partial \bfe_{k}}{\partial
s_{\alpha}}\right) \left(\Delta\bR^{\rm T}\frac{\partial
\bfe_{k'}}{\partial s_{\alpha}}\right). 
\eea 
Based on above
equations we can simplify  \Eq{ke1} to yield (cf. Ref. \citenum{miller88:6298})
\bea\label{kelinear}
T\left(\bs,\bQ\right)&=&\frac{1}{2}\sum_{\alpha}P_{\alpha}^2
+\frac{1}{2}\sum_{kk'}P_{k}^{\dag}
\left(\delta_{kk'}+\sum_{\alpha}B_{\alpha k}B_{\alpha
k'}\right)P_{k'} \nl &&+\left(\frac{1}{2}P_{\alpha}\sum_{\alpha
k}B_{\alpha k}P_k+{\rm h.c.}\right), 
\eea 
where $B_{\alpha k}=\Delta\bR^{\rm T}\partial \bfe_{k}/\partial s_{\alpha}$. 
Note here all the components of
momentum are Hermitian according to Ref. \citenum{yang08:2445}. The
kinetic couplings are caused by the $\bs$-dependence of $\{\bfe_k\}$
as can be seen from the expression for  $B_{\alpha k}$. The potential
energy is still given by \Eq{pot2}. 

The KEO can be further simplified by using more constant unit vectors for the expansion of the coordinate space, i.e., we get rid of the
$\bs$-dependence of $\{\bfe_k\}$ (for alternative approaches see also Refs. \citenum{miller88:6298,yang08:2445}). The most simple case, in which the
kinetic energy has a quite trivial form while the potential energy
is no longer diagonal, is the space whose unit vectors are all
constants. This can be achieved by diagonalizing the projected
Hessian matrix at only one point, $\bR_{\rm ref}$, instead of  each
point on the reaction surface. The new representation is obtained by
a pure $\bs$ independent rotation and the new variables are defined
by 
\bea 
s_{\alpha}=\bfe_{\alpha}^{\rm T}\left(\bR-\bR_{\rm ref}\right) \nl
Q_k=\bfe_{k}^{\rm T}\left(\bR-\bR_{\rm ref}\right). 
\eea 
Here \{$Q_k$\} denote the remaining $3N-D$ variables which are the global translation,
rotation and normal modes only at the reference point. Notice, that within this approximation overall rotations are not strictly projected out
for a general point on the potential energy surface. 
The Hamiltonian in terms of the new coordinates reads 
\bea\label{ke3} 
T\left(\bs,\bQ\right)
&=&\frac{1}{2}\sum_{\alpha}P_{\alpha}^2+\frac{1}{2}\sum_{k}P_{k}^2
\nl &=&-\frac{1}{2}\sum_{\alpha}\frac{\partial^2}{\partial
s_{\alpha}^2} -\frac{1}{2}\sum_{k}\frac{\partial^2}{\partial
Q_{k}^2} \nl V\left(\bs,\bQ\right)&=&V\left(\bRz\right) -\sum_k
f_k\left(\bs\right) Q_k +\frac{1}{2} \sum_{k,k'}
K_{kk'}\left(\bs\right) Q_kQ_{k'}, 
\eea
where $f_k$ has the same definition as before and
\begin{equation}
\label{eq:k}
K_{kk'}\left(\bs\right)=\bfe_k^{\rm T} \frac{\partial^2 V}
{\partial \bR ^2}\Big|_{\bRz} \bfe_{k'} \,.
\end{equation}
This form of the Hamiltonian has been used in the present paper to model the coupling between the reaction coordinate and the intramolecular vibrational modes in the application to the proton transfer in 6-Aminofulvene-1-aldimine.
\clearpage



%
\end{document}